\documentclass{ieeeaccess}
\usepackage{amsmath,amssymb,amsfonts}
\usepackage{algorithmic}
\usepackage{graphicx}
\usepackage{textcomp}
\usepackage{array}
\usepackage[utf8]{inputenc} 
\usepackage[T1]{fontenc}    
\usepackage{amssymb}
\usepackage[graphicx]{realboxes}
\usepackage{url}            
\usepackage{booktabs}       
\usepackage{nicefrac}       
\usepackage{microtype}      
\usepackage{amsmath}
\usepackage{multirow}
\usepackage{graphicx}
\usepackage[numbers]{natbib}
\usepackage{caption}
\usepackage{bm}
\usepackage{hyperref}     
\usepackage{cleveref}      

\makeatletter
\AtBeginDocument{\DeclareMathVersion{bold}
\SetSymbolFont{operators}{bold}{T1}{times}{b}{n}
\SetSymbolFont{NewLetters}{bold}{T1}{times}{b}{it}
\SetMathAlphabet{\mathrm}{bold}{T1}{times}{b}{n}
\SetMathAlphabet{\mathit}{bold}{T1}{times}{b}{it}
\SetMathAlphabet{\mathbf}{bold}{T1}{times}{b}{n}
\SetMathAlphabet{\mathtt}{bold}{OT1}{pcr}{b}{n}
\SetSymbolFont{symbols}{bold}{OMS}{cmsy}{b}{n}
\renewcommand\boldmath{\@nomath\boldmath\mathversion{bold}}}
\makeatother

\def\BibTeX{{\rm B\kern-.05em{\sc i\kern-.025em b}\kern-.08em
    T\kern-.1667em\lower.7ex\hbox{E}\kern-.125emX}}

\begin{document}
\history{This article has been accepted for publication in \href{http://dx.doi.org/10.1109/ACCESS.2025.3648056}{IEEE Access}. Content may differ from the final edited publication.}
\doi{ 10.1109/ACCESS.2025.3648056}

\title{Benchmarking Time Series Foundation Models for Short-Term Household Electricity Load Forecasting}
\author{\uppercase{Marcel Meyer}\authorrefmark{1},
\uppercase{David Zapata Gonzalez}\authorrefmark{1},\uppercase{Sascha Kaltenpoth}\authorrefmark{1},\uppercase{Oliver Müller}\authorrefmark{1}}

\address[1]{Data Analytics Group, Paderborn University, 33098 Paderborn, Germany}
\tfootnote{This study was supported by the Federal Ministry for Environment,
Nature Conservation, and Nuclear Safety of Germany under Grant
No. 67KI32008B (DC2HEAT - Data center HEat Recovery with AI-
Technologies) and the Federal Ministry for Economic Affairs and Energy under Grant No. 03EN6024A (DynOpt-San), and we gratefully acknowledge their support.}

\markboth
{Meyer \headeretal: Benchmarking Time Series Foundation Models for Short-Term Household Electricity Load Forecasting}
{Meyer \headeretal: Benchmarking Time Series Foundation Models for Short-Term Household Electricity Load Forecasting}

\corresp{Corresponding author: Marcel Meyer (e-mail: marcel.meyer@uni-paderborn.de).}

\begin{abstract}
  Accurate household electricity short-term load forecasting (STLF) is key to future and sustainable energy systems. While various studies have analyzed statistical, machine learning, or deep learning approaches for household electricity STLF, recently proposed time series foundation models such as Chronos, TimesFM or Time-MoE promise a new approach for household electricity STLF. These models are trained on a vast amount of time series data and are able to forecast time series without explicit task-specific training (zero-shot learning). In this study, we benchmark the forecasting capabilities of time series foundation models compared to Trained-from-Scratch (TFS) Transformer-based approaches. Our results suggest that foundation models perform comparably to TFS Transformer models, while certain time series foundation models outperform all TFS models when the input size increases. At the same time, they require less effort, as they need no domain-specific training and only limited contextual data for inference.
\end{abstract}


\titlepgskip=-21pt

\maketitle
\section{Introduction}
\label{introduction}

The energy transition, especially the incorporation of renewable energy sources into our energy system, leads to increased electricity load variability, as more households act simultaneously as generators and consumers \cite{tavakoli_impacts_2020}, electric vehicles introduce additional irregular load into the grid \cite{Abid.2024} and the electrical grid is decentralized into micro-grids \cite{Shaukat.2023}. From a distribution system operator's (DSO's) perspective, forecasting the energy consumption of private households poses unique challenges, as their load profiles depend on various (unobserved) factors like household size, installed appliances, or own energy generation (e.g., PV). Therefore, households are often black boxes, leaving reasons for consumption variability unrevealed \cite{Paatero.2006}. Additionally, the sheer number of private households leads to massive data management and processing challenges, which require sophisticated machine learning pipelines with continuous training and evaluation of models. Consequently, an efficient and effective electricity demand prediction based on diverse, univariate load time series in the short term, particularly at the low-voltage household level, is vital to ensure a resilient and intelligent electricity distribution.

Various studies have compared univariate low-voltage, household, or residential electricity short-term load forecasting (STLF) approaches \cite{Hertel.2023, Mathumitha.2024, Vanting.2021}, providing evidence that univariate deep learning approaches based on the Transformer architecture \cite{Vaswani.2017} outperform other univariate approaches \cite{Hertel.2023, Upadhyay.2023}. While Trained-from-Scratch (TFS) Transformers models deliver accurate and fast forecasts, it is necessary to train them from scratch for every specific domain or task (e.g., type of household or geography) and retrain them in regular intervals (e.g., every season).

The recent development of time series foundation models (TSFM), which are (pre-)trained on large and diverse time series datasets, offers the possibility to depart from the traditional method of training one model per task and iteratively retraining it.
Out-of-the-box, without further domain adaptation or fine-tuning \cite{ruder2019neural}, these models can accurately (zero-shot) predict univariate time series from historical data \cite{Liang.2024}. This advancement could transform the way we forecast household electricity loads by enabling straightforward predictions without continual and task-specific retraining. However, whether massive pre-training of Transformers on very large collections of generic time series (e.g., Finance, Healthcare, Traffic, Energy) \cite{Ansari.2024, Rasul.2024} can actually represent household load patterns in real-world scenarios is an empirical question that so far has not been answered, especially considering new evaluation hurdles coming alongside these global models. Accordingly, this study is guided by the central research question:

``Can zero-shot TSFMs match the capabilities of state-of-the-art trained-from-scratch Transformers in forecasting household electricity load?''

To avoid overestimating the performance of TSFMs, it is particularly important that the evaluation data is not already included in the pre-training data. Consequently, in this study, we compare existing state-of-the-art (SOTA) TFS Transformer forecasting models (trained on household electricity time series) with TSFM to determine the suitability of foundation models in household electricity STLF.

We evaluate in our benchmark with two real-world datasets from Germany and two real-world datasets from Great Britain, leading in total to over 300 individual households. All datasets reflect a realistic use case from a DSO's perspective, including households with different start and end times, several load profiles, and seasonal fluctuations.

Our results suggest that TSFM are comparable to TFS Transformers in terms of accuracy. We found that depending on the metric, Time-MoE \cite{shi_time-moe_2024}, Sundial \cite{liu2025sundial}, Chronos \cite{Ansari.2024} and TimesFM \cite{Das2024} provide competitive forecasting capabilities. Especially on longer input sizes, TSFM outperformed multiple recent TFS Transformer approaches without being fine-tuned on the task. This finding suggests that with domain adaptation (e.g., creating a foundation model for diverse energy load forecasting tasks) or fine-tuning (e.g., on historical data from the time series under consideration), time series foundation models might become a promising research direction in household electricity STLF.

The remainder of this paper is structured as follows. First, we summarize related work on univariate household electricity STLF and briefly explain the theory behind foundation models. Next, we describe the methodology of our comparative benchmarking study in detail. Subsequently, we show and discuss the empirical results of our experiments. Lastly, we provide a conclusion and outlook for future research on energy-related time series foundation models.

\section{Related Work}
\label{section:related-work}

Statistical approaches (e.g., SeasonalAverage, ARIMA), as well as more recently machine learning and deep learning approaches based on neural networks, dominate the field of low-voltage level electricity STLF \cite{Tzafestas2001,Haben.2021,Tarmanini.2023, Vanting.2021}. \citet{Hopf2023} conducted a meta-analysis on household electricity STLF and showed that especially hybrid neural networks (NNs) and long short-term memory (LSTM) NNs significantly reduce the forecasting error on the individual (i.e., low-voltage household) level.

Considering deep learning approaches, recent studies show that forecasting methods based on the Transformer \cite{Vaswani.2017} architecture tend to outperform other approaches, especially LSTMs, in diverse forecasting domains \cite{Ahmed2023, Lara.2021, Li2023, Nascimento2023, Sun2023}. Furthermore, various studies compared the performance of different variants of Transformer architectures. \citet{wen_Transformers_survery_2022} compare different Transformer architectures' forecasting accuracy using different input lengths and different numbers of layers. While they could not determine a superior architecture, they found that using input sizes exceeding the horizon decreases the forecasting performance of Transformer-based architectures, whereas the performance increases with a rising number of layers \cite{wen_Transformers_survery_2022}. In contrast \cite{GAO2023100142} found that combining different learning strategies in an adaptive theory-guided framework improves performance compared to the vanilla transformer.

Returning to electricity STLF, many studies focused on electricity STLF at the substation \cite{Giacomazzi.2023} or grid level \cite{Zhang.2022, Zhao.2021}, or investigate multivariate methods \cite{Zhang.2021, Cen.2024}, while only four studies focus on the use of TFS Transformers in univariate household electricity STLF (see Table \ref{tab:related_work_studies}).

\begin{table*}[ht]
\centering
\scriptsize
    \caption{Recent studies on univariate household electricity STLF}
    \begin{tabular}{ccccclll}
    \hline \hline
        Multiple & Baseline & Cross- & Transformer & Foundation & Best Models & Source \\
        Datasets & ~ & Validation & ~ & Model & ~ & ~ \\
    \hline \hline
        ~ & X & X & X & ~ & VanillaTransformer$^{1}$ & \cite{Upadhyay.2023} \\ 
        ~ & X & X & X & ~ & VanillaTransformer & \cite{Sievers.2023} \\ 
        ~ & X & X & X & ~ & PatchTST$^{2}$ & \cite{Cen.2024} \\ 
        X & X & X & X & ~ & PatchTST & \cite{Hertel.2023} \\ 
        X & X & X & X & ~ & VanillaTransformer & \cite{emami_buildingsbench_2024} \\
        X & X & X & X & X & VanillaTransformer$^{3}$ & \cite{saravanan_analyzing_2024} \\
        X & X & X & X & X & see Section \ref{sec:results} & ours \\
    \hline \hline
    \end{tabular}
    {\\ $^{1}$Modification in training strategy $^{2}$Modification in embedding structure \par $^{3}$Information leakage in evaluation \par}
    \label{tab:related_work_studies}
\end{table*}

\citet{Upadhyay.2023} proposed a VanillaTransformer with a modified training strategy, which predicts the 25th hour based on the historical 24 hours, while the loss is only computed for the 25th forecast value. They compare their approach with diverse machine learning and deep learning algorithms, including random forests, CNN, and LSTM architectures. Furthermore, they show that the Transformer performs best, directly followed by LSTMs. 

The study of \citet{Sievers.2023} compared local, central, and federated learning for CNN, LSTM, and Transformer models. They found that the VanillaTransformer performed best in every training scenario and that local learning, where one model is trained on every dataset, and federated learning, where models are trained locally and then merged into a global model, perform equally well. In contrast, the central learning strategy, where one model is trained using all data on a central server, performed worst \cite{Sievers.2023}. 

\citet{Cen.2024} developed a modified PatchTST \cite{nie2023time} and compared it to a GRU, LSTM, and multiple other Transformer variants. Their modified PatchTST model outperformed all other approaches, followed by the VanillaTransformer.

While all recent studies incorporate baselines and times series cross-validation, only the study of \citet{Hertel.2023} used multiple datasets. They compared diverse Transformer-based approaches with linear regression, an ANN, and an LSTM for forecasting hourly values with a 24-hour, 96-hour, and 720-hour horizon on two datasets. Additionally, they investigated three different training strategies: (1) a local training strategy that trains a separate model for every household, (2) a multivariate strategy that trains one model to predict all households at the same time, and (3) a global strategy that trains one model to forecast multiple households but only one at the same time. In their study, a globally trained PatchTST \cite{nie2023time} model performed best for a 96-hour and 720-hour horizon, and a globally trained VanillaTransformer \cite{Vaswani.2017} performed best for a 24-hour horizon.

The comparison of the related work summarized in Table \ref{tab:related_work_studies} suggests that the VanillaTransformer and PatchTST represent the current SOTA in TFS Transformer architectures. Furthermore, two additional TFS Transformer-based architectures seem promising: The recently published iTransformer model \cite{Liu.2024} and the Temporal Fusion Transformer (TFT), which is based on a mixture of LSTM and the attention mechanism \cite{lim2021temporal} and has shown competitive performance in substation electricity STLF tasks \cite{Giacomazzi.2023}.

Considering the positive results of recent studies on global forecasting models (\cite{Hertel.2023}), the approach to train Transformer-based models on vast amounts of time series data comprising different domains and frequencies as so-called foundation models seems to be a promising avenue for future research on STLF. As general-purpose zero-shot forecasting models, these pre-trained models are able to accurately predict time series without fine-tuning or retraining them on the domain or task-specific datasets \cite{Liang.2024}. 

Two types of foundation models can be distinguished. Large language model-based foundation models, which convert time series into textual representations, sometimes enhanced by other architectures such as Graph Neural Networks, such as PromptCast \cite{Xue.2023}, LLMTime \cite{Gruver.2023}, and FSCA \cite{hu2025contextalignment} represent the first type of time series foundation models. However, these models have a high resource utilization and lack scalability and practicability \cite{Ansari.2024,Tan2024}. Transformer-based architectures for time series such as TimeGPT-1 \cite{Garza.2023}, LagLlama \cite{Rasul.2024}, Chronos \cite{Ansari.2024}, and TimesFM \cite{Das2024} comprise the second type of foundation models. Inspired by large language models (LLM), these models are specifically trained on tokenized time series to forecast the most probable token, which encodes an explicit part of a time series.

While these models showed impressive zero-shot capabilities in various domains \cite{aksu_gift-eval_2024}, their vast training datasets create a unique evaluation problem. Specifically, many publicly available benchmarking datasets have been used to train these foundation models. Therefore, it is crucial to evaluate them on out-of-sample datasets not included in the foundation models' training data.
In fact, it can quickly happen that the selected evaluation data is already available in the sheer volume of training data. For example, a study evaluated TSFM on the Buildingsbench Dataset \cite{emami_buildingsbench_2024}, which at first glance appears suitable for an STLF evaluation \cite{saravanan_analyzing_2024}. However, the Buildingsbench Dataset partially bundles other datasets that are already included in many of the TSFM training data, such as the London Smart Meters Dataset (Chronos, Moirai, Time-MoE) or the Portuguese Household Dataset (Chronos, TimesFM, Moirai, Time-MoE). Since information leakage is a potential issue here, the Buildingsbench Dataset is not suitable as an TSFM evaluation dataset for STLF.

As Table \ref{tab:related_work_foundation_models} summarizes, most TSFM provide information on their (pre-)training data, which enables an evaluation without test set contamination. In contrast, information about the training data for TimeGPT-1 is not publicly available, disqualifying it from an appropriate evaluation using historical open-source datasets.

\begin{table}[h]
    \centering
    \caption{Time series foundation model training data disclosure}
    \begin{tabular}{lcl}
    \hline
    \hline
        Model & Open data & Source \\
    \hline
    \hline
        Chronos(-Bolt) & X & \cite{Ansari.2024} \\
        LagLlama & X & \cite{Rasul.2024} \\
        Moirai(-MoE) & X & \cite{woo_unified_2024, liu_moirai-moe_2024}  \\
        Time-MoE & X & \cite{shi_time-moe_2024}\\
        Sundial & X & \cite{liu2025sundial}\\
        TimeGPT-1 &   & \cite{Garza.2023} \\
        TimesFM (2.0)  & X & \cite{Das2024} \\
    \hline
    \hline
    \end{tabular}
    \label{tab:related_work_foundation_models}
\end{table}

It is important to note that the typical evaluation strategy used in TSFM often differs from standard time series cross-validation. Instead of applying a rolling-window or expanding-window validation over the entire time series, TSFMs are usually assessed only on the final observations of the series that corresponds to the forecast horizon \cite{Das2024, Ansari.2024, liu2025sundial}. In other words, the evaluation does not track how the model performs over time but is instead based on producing a single prediction for the last n time points, with performance aggregated across the collection of time series rather than across multiple time segments.

\section{Method and Data}
STLF approaches at the industry-, building-, and household-levels have been investigated in diverse studies \cite{Haben.2021}. While \citet{Haben.2021} did not identify an approach that is superior in all situations in their review, they emphasized three important factors for the evaluation of low-voltage electricity STLF approaches: (1) The evaluation must be applied on multiple datasets, (2) it should include appropriate naive and sophisticated baselines, such as SeasonalAverage or deep learning models, and (3) it should apply time series cross-validation.
Hence, we developed a data acquisition, model selection, and evaluation strategy that fulfills these requirements.

\subsection{Data Acquisition}
\label{section:data-acquisition}

\begin{table*}[ht]
\centering
\caption{Household Energy Datasets used in foundation model pre-training} \label{tab:used_datasets}
\begin{tabular}{lcccccc}

\hline\hline
Datasets & Chronos (-Bolt) & LagLlama & Moirai (-MoE) & Sundial & Time-MoE & TimesFM (2.0) \\
\hline\hline
Ausgrid Solar Home Dataset      &   & X & X & X & X & \\
\hline
REFIT Dataset*          &   &   &   &   &    &   \\
\hline
Electricity Dataset             & X & X &   & X & X & X \\
\hline
London Smart Meters Dataset     & X & X & X & X & X & \\
\hline
IDEAL Dataset*                &   &   & X & X & X &  \\
\hline
Lower Saxony Dataset*           &   &   &   &   &    &  \\
\hline
Southern Germany Dataset*       &   &   &   &   &    &  \\
\hline\hline
\end{tabular}%
\vspace{0.5em}
\end{table*}

In Table \ref{tab:used_datasets}, we show which datasets are used in the pre-training of the TSFM. We first considered a total of nine datasets for benchmarking:  The \textbf{Electricity Dataset} \cite{Trindade2015} with hourly electricity consumption from households, shops, and industrial business in Portugal, the \textbf{Ausgrid Solar Home Dataset} \cite{ratnam_2017} with solar energy production and private consumption from clients in Australia, and the \textbf{London Smart Meters Dataset} \cite{UKPower-2015} from electrical consumption of Households in the United Kingdom are all included in the training set of multiple TSFM.

Therefore, they cannot be used for evaluation without the risk of leakage \cite{kapoor_leakage_2022} and an overestimation of the performance of the foundation model. The \textbf{French Household Dataset} \cite{misc_individual_household_electric_power_consumption_235} contains only a single house, lacking diversity, and the \textbf{Danish Dataset} \cite{energinet-2021} contains energy consumption of whole districts, which is not the focus of our study. Three sourced datasets are not included in any pre-training and can be used for evaluation without any restriction: The \textbf{Southern Germany Dataset} \cite{OPSD_2020}, the \textbf{Lower Saxony Dataset} \cite{Schlemminger2022} and the \textbf{REFIT Dataset} \cite{Murray2017}.

The \textbf{REFIT Dataset} contains household data from the Loughborough area in the United Kingdom \cite{Murray2017} with house characteristics and appliance-by-appliance energy consumption per minute for two years. In this study, we used the hourly aggregated consumption for each of the 20 households.

The \textbf{Southern Germany Dataset} comprises electricity consumption from small businesses and households in the city of Konstanz in Germany \cite{OPSD_2020}. We filtered the data for households only. Some houses also generate energy from PV panels, and some show consumption from individual devices, such as dishwashers, freezers, heat pumps, and, for one house, an electric vehicle. Keeping the perspective of the DSO, we filtered the data to obtain the grid's total electricity import from the six households in the dataset. Although this dataset contains only six households, its primary value lies in the extensive duration of data collection rather than the number of unique entities. Since our approach involves the training of TFS Transformers and the application of time-series cross-validation, the temporal depth is the deciding factor for including this dataset. This extended duration allows us to incorporate multiple seasonality patterns into the training while simultaneously yielding a robust test set of 16,959 observations.

The \textbf{Lower Saxony Dataset} holds electrical single-family house consumption and partly PV energy generation near the city of Hameln in Germany \cite{Schlemminger2022}. We selected the \textit{active power} for all measured phases. The dataset also includes pumps that are used in the district heating network. These are measured via a separate smart meter and can be treated separately by the DSO and, hence, are not considered in the analysis. Furthermore, four households with PV systems were excluded. Their metering configuration did not distinguish clearly between grid import and PV generation (net metering), resulting in ambiguous net-load profiles containing negative values. To ensure a consistent target variable representing household demand, these time series were deemed unsuitable for the evaluation and removed.

The \textbf{IDEAL Dataset} \cite{ideal_2021} represents a special case: on the one hand, it contains a large number of households, which significantly increases the informative value of the evaluation. On the other hand, it was used in the Moirai, Sundial and Time-MoE training data. We decided to include the dataset in the evaluation and to exclude the evaluation of the three TSFMs for this dataset. Details can be found in section \ref{section:metrics}. The \textbf{IDEAL Dataset} covers electric, gas, temperature, humidity, and metadata from households in Edinburgh and nearby regions in the UK \cite{ideal_2021}. Some houses contain information about electrical appliances and more detailed information about temperature and gas and heating equipment, as well as weather information. In the study, we use the net load electricity consumption from each of the 254 households.
\begin{table*}[ht]
\centering
    \centering
    \caption{Information about the datasets}
    \begin{tabular}{>{\raggedright\arraybackslash}p{2cm}  
                    >{\raggedright\arraybackslash}p{2cm}  
                    >{\raggedright\arraybackslash}p{2cm}  
                    >{\raggedright\arraybackslash}p{0.7cm}  
                    >{\raggedright\arraybackslash}p{1.3cm}  
                    >{\raggedright\arraybackslash}p{1.3cm}  
                    >{\raggedright\arraybackslash}p{1.3cm}}   
    \hline
    \hline
    Dataset & Start Date & End Date &  Nr. & Mean & Median & Std \\
      \hline
      \hline
        IDEAL & 10/08/2016 & 01/07/2018 & 254 & 0.3713 & 0.2220 & 0.4395 \\
        \hline
        Lower Saxony & 02/05/2018 & 31/12/2020 & 34 & 0.3417 & 0.2337 & 0.3411  \\
        \hline
        Southern Germany & 15/04/2015 & 06/09/2017 & 6 & 0.4035 & 0.2900 & 2.6643  \\
        \hline
        REFIT & 17/09/2013 & 10/07/2015 & 20 & 0.5151 & 0.3279 & 0.5197 \\
      \hline
      \hline
    \end{tabular}
    \label{tab:dataset_info}
\end{table*}
\begin{figure}[h]
  \centering
  \includegraphics[width=0.9\linewidth]{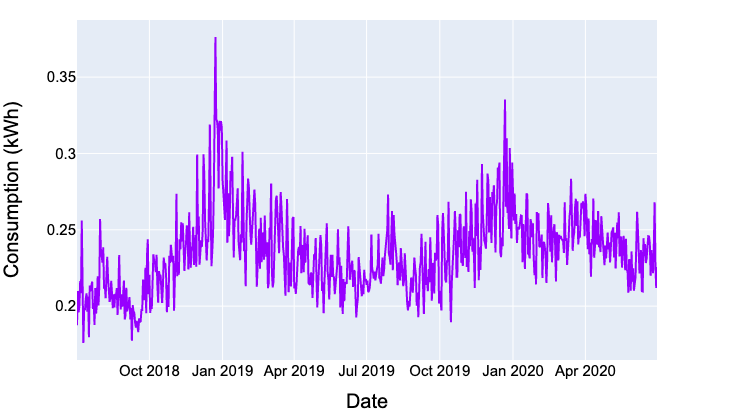}
  \caption{Median hourly energy consumption per day for all houses in the Lower Saxony dataset}
  \label{fig:eda_5_1_timeserie_1.png} 
\end{figure}

\begin{figure}[h]
  \centering
  \includegraphics[width=0.8\linewidth]{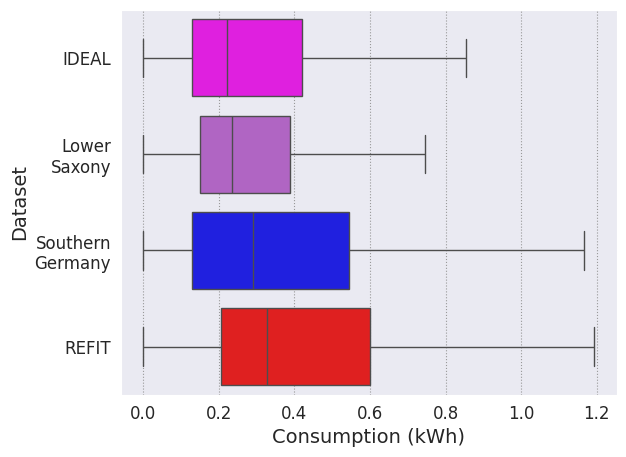}
  \caption{Distribution of energy consumption per hour}
  \label{fig:eda_6_boxplot.png} 
\end{figure}

For all datasets, we used hourly aggregated measurements. The unit of electricity consumption is expressed in kilowatt-hours (kWh). The general information about the evaluation datasets is summarized in Table \ref{tab:dataset_info}.

\subsection{Exploratory Data Analysis (EDA)}

Figure \ref{fig:eda_5_1_timeserie_1.png} shows the median hourly daily consumption for all Lower Saxony households, revealing a strong seasonal pattern with a peak energy consumption in January. The IDEAL, Southern Germany, and REFIT datasets show similar patterns. All time series are non-stationary. Southern Germany has from Juli 2017 on the same constant values every day, so we dropped this part of the data.

Figure \ref{fig:eda_6_boxplot.png} illustrates the distribution of energy consumption for the four datasets. The box plots suggest a distribution with positive skewness. Lower Saxony and IDEAL depict a distribution with a positive kurtosis, while Southern Germany and REFIT have flatter distributions with long positive tails and more outliers.

\begin{figure}[h]
  \centering
  \includegraphics[width=0.9\linewidth]{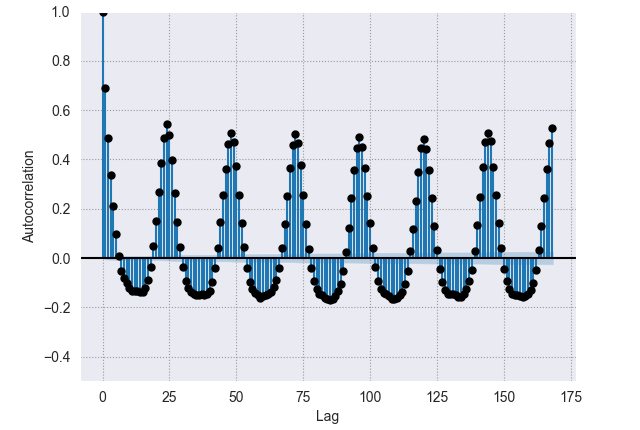}
  \caption{Autocorrelation with 168 lags for "residential house 3" in Southern Germany}
  \label{fig:eda_3_autocorrelation.png} 
\end{figure}

\begin{figure}[h]
  \centering
  \includegraphics[width=0.9\linewidth]{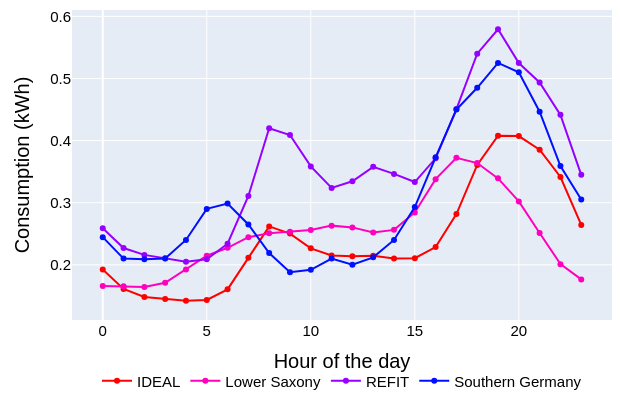}
  \caption{Median hourly energy consumption}
  \label{fig:eda_2_timeserie.png} 
\end{figure}

\begin{figure}[h]
  \centering
  \includegraphics[width=0.9\linewidth]{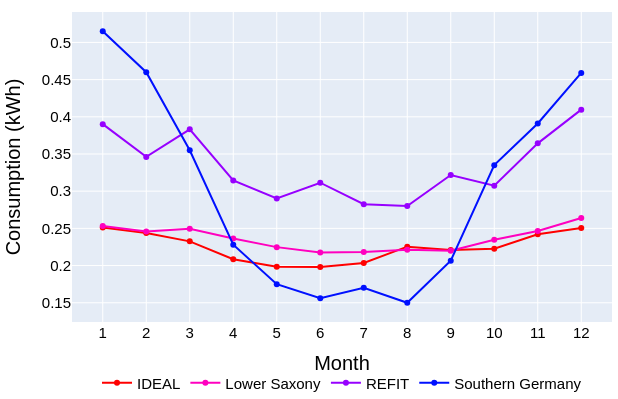}
  \caption{Median hourly energy consumption per month}
  \label{fig:eda_4_timeserie.png} 
\end{figure}

Figure \ref{fig:eda_3_autocorrelation.png} shows the autocorrelation for a single house's median electricity load per hour in Southern Germany during March 2016. A cyclical pattern is visible in the time series in a sinuous form, also described in \cite{Mouakher2022} for 24-hour lags, where the energy consumption for an hour is strongly correlated with the same hour of the following day. Daily patterns and weekly patterns are common for many houses in all datasets.

Figure \ref{fig:eda_2_timeserie.png} depicts the median consumption per hour over all days for each dataset. The time series show an increase in electricity usage during the morning hours, a slight increase in Lower Saxony and a decrease in the rest in the afternoon, a strong rise in the evening hours for all datasets, and a sharp fall during the night hours. 

Figure \ref{fig:eda_4_timeserie.png} shows the median consumption per hour in the different months of the year. There is a clear reduction in consumption from spring to fall in comparison to the winter months. This change is very apparent in the Southern Germany and REFIT datasets, and less visible in the Lower Saxony and IDEAL dataset.

It is important to consider that the datasets contain distinct numbers of households and different household characteristics, such as size,  number of occupants, energy needs, the presence or absence of PV installations, and sometimes the use of heat pumps for heating. This last point could explain why the consumption in some datasets differs from the rest and is more nuanced by the seasons. The focus of this study is not to explain the causes of the differences in the datasets, but to use them to compare the models. Such heterogeneous datasets represent a realistic view of energy consumption modeling challenges, where the DSOs have no insights into the energy demand causes in individual households.

The datasets have individual missing values between two measurements or, in the case of the Lower Saxony and REFIT datasets, up to weeks-long gaps. The reasons given in the datasets' descriptions are technical failures of the data logger, Internet outages, conversion work \cite{Schlemminger2022}, failed radio transmissions, and problems in daylight savings time transitions \cite{OPSD_2020}. 

Several data exploration results need to be considered for the preprocessing and modeling:
\begin{itemize}
\item the time series have seasonal and cyclical components and are non-stationary.
\item the time series show a similar cyclical pattern, with high autocorrelation in 24-hour lags.
\item the datasets represent different data distributions.
\item the Lower Saxony and REFIT dataset has long gaps in the measurements.
\item the household measurements start and end at different times.
\end{itemize}

\subsection{Data Preprocessing}
\subsubsection{Handling Missing Values}
To ensure a complete time series for each household, we generated entries for all hours between the start and end of each time series, filling non-existing entries with missing value representations.
Most datasets have multiple days or weeks with missing values. These gaps were not interpolated, as households can develop dynamically over several days. Instead, we have taken the longest possible time period in which no interruption lasted longer than three consecutive days.

We carried out linear interpolation for the remaining missing values embedded within the time series. If the missing values occur at the end of the series, we took the value from 24 hours ago due to the strong autocorrelation in the datasets.

We only handled the missing values in the training data, and intentionally didn’t touch them in the evaluation set. We did this to make sure that we are validating with real measurements.

\subsubsection{Train-Test Split}
\begin{figure*}[h!]
  \centering
  \includegraphics[width=1\linewidth]{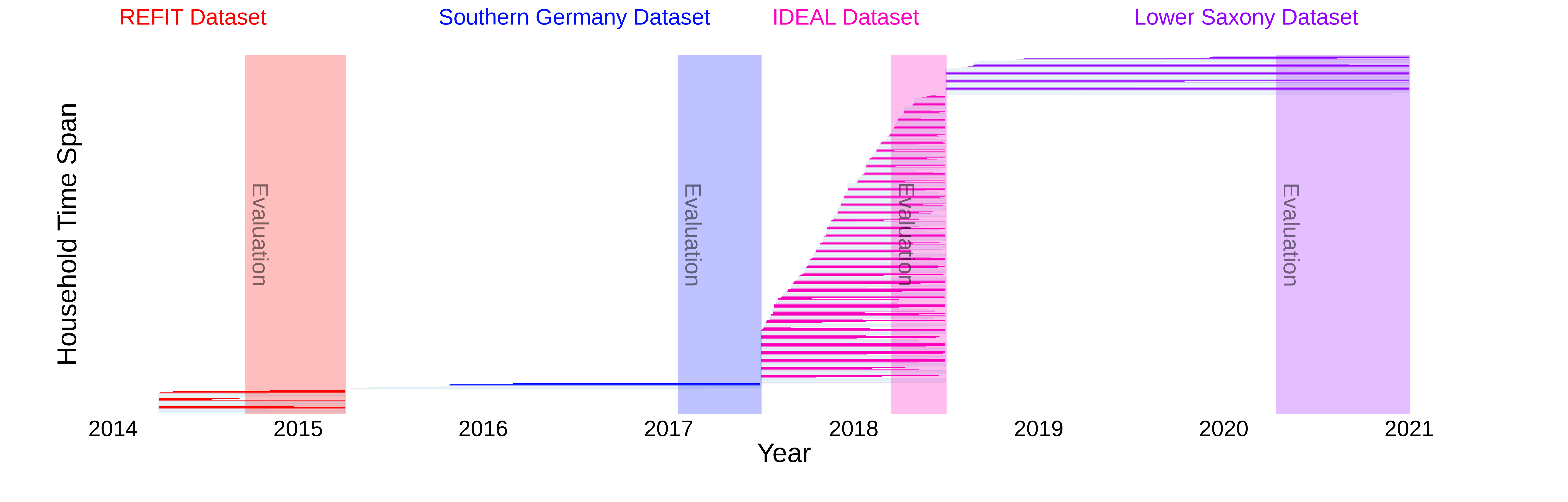}
  \caption{Train-Test split. Time spans of each time series}
  \label{fig:train_test_split.png} 
\end{figure*}
Figure \ref{fig:train_test_split.png} illustrates the start and end dates of the measurements taken from the datasets. A good proportion of the households have no common start and/or end date. This represents realistic energy data from households, with new houses connected to the grid and others disconnected, rather than a clean dataset where all houses start and end on the same dates. However, this poses a challenge for the training and evaluation of models.

Using the train-test method proposed by \citet{Hertel.2023}, splitting the data into 70 \% training, 10 \% validation, and 20 \% testing would result in different split dates for each household in our datasets. This presents a risk of information leakage, for instance, global time-specific patterns, such as the Covid-19 crisis, could be represented in one time series training data (Household A). Another time series (Household B) could have a different split date with a test set that comprises the same time period as the training data of other households (Household A). When using this approach, the global pattern might be learned (training data Household A) by global models and transferred to other time series (Household B). To safeguard against the leakage of information, we select a unique split-date for each dataset.
Additionally, we ensure no overlapping of the datasets by cutting them if the evaluation time frame of one dataset overlaps with the training data of another dataset.

In order to find a balance between the amount of test data (last x percentage of data) and evaluating the maximum number of households in an overlapping time frame, we implemented the following logic for defining the split date. 

First, we determine the time point $t_{0.25}$ of the 0.25th percentile of each household's maximum date (Step 1). In Step 2, we calculate all possible time points $T_{poss}=(t_{min}, ..., t_{0.25})$ (i.e., hours) between the global minimum date $t_{min}$ and $t_{0.25}$, regardless of their frequency in the actual data. Based on this, we determine the final train-test split at the 0.8th percentile of $T_{poss}$ (Step 3). By ensuring that the date defined in Step 1 is included in our test set, we ensure that the test set comprises a large number of households.

The split parameters stay the same for each dataset except for the widely spread Southern Germany dataset, where a percentile of 0.5 of maximum dates (Step 2) leads to a split date containing 100 \% of all households having data on that day and a test size of around 19 \% of all data.

\subsection{Evaluated Models} \label{section:selected-models}

A condition for the selection of TSFM was the disclosure of the training data and the provision of a pre-trained model as discussed in chapter \ref{section:data-acquisition}. At the time of writing, Chronos \cite{Ansari.2024}, TimesFM \cite{Das2024}, Sundial \cite{liu2025sundial}, Time-MoE \cite{shi_time-moe_2024}, Moirai \cite{woo_unified_2024} and LagLlama \cite{Rasul.2024} fulfill both conditions. The TSFM are not be fine-tuned on the household datasets introduced for our evaluation. Only inference is used on the unseen data for achieving zero-shot predictions \cite{zero-shot-xian2019}. Therefore we excluded models which need finetuning e.g. the TSFM Moment \citet{goswami_moment_2024}, which forecasting head is randomly initialized. The implemention is done by following the suggested usage from the official Github repository and using the provided model weights. For LagLlama, we activate RoPe scaling as suggested in the zero-shot tutorial for input sizes longer than the context length of 32, because most of the tested input sizes are above that threshold.
As the main TFS competitors for the foundation models, we used time series models with a similar Transformer-based architecture. Namely, we chose the original Transformer Encoder-Decoder implementation ("VanillaTransformer") \cite{Vaswani.2017}, and PatchTST \cite{nie2023time}, which represents the current SOTA for household energy STLF (see Table \ref{tab:related_work_studies}). Furthermore, we chose iTransfomer \cite{Liu.2024} and Temporal Fusion Transformer (TFT) \cite{lim2021temporal}, which we identified in our literature review. The TFS Transformers are trained univariate, which can be a limiting factor for the TFS Transformers to varying degrees. However, this approach allows for the most direct comparison with the foundation models.
For implementing the TFS Transformers, we used the \href{https://github.com/Nixtla/neuralforecast}{NeuralForecast library}, which is referring its model implementations to the corresponding papers.
\footnote{The complete preprocessing, training and evaluation code can be found under this github repo: \url{https://github.com/mmcux/benchmarking_tsfm_household_load_forecasting}}

As a baseline, we take the SeasonalAverage \cite{hyndman2018forecasting} with a seasonality of 24 as we observed a high autocorrelation in the data with lags of 24. The SeasonalAverage is calculated per time series with the Pandas library with an hourly average over the input size as forecasts for the horizon.
Additionally, we report as baseline reference the Naive Forecast where the last observed value is the forecast for the complete forecast horizon.

This study's primary objective is to assess the different approaches' general ability to adapt to new data. We vary the input size for all models to identify the influence of data context for pattern recognition. We also refrain from extensive hyperparameter tuning. We gathered over 2 million training data points, which, in combination with compute-intensive TFS Transformer training, would require a substantial amount of compute resources for hyperparameter tuning. Thus, we use the default parameters of the models as proposed in their implementation, wich are usually determined as optimal for large datasets in the corresponding papers.

A complete overview of the used models and hyperlinks to their implementation can be found in Table \ref{tab:model_architectures}.

\begin{table*}[ht]
\centering
\captionof{table}{Comparison of models, their architectures and implementations}
\begin{tabular}{lcccccc}
\hline
\hline
\multicolumn{1}{c}{Model Name} & \multicolumn{1}{c}{Model Architecture} & \multicolumn{1}{c}{Model Type} & \multicolumn{1}{c}{Implementation} \\
\hline
\hline
Chronos & Transformer-based & Foundation Model (F) & \href{https://github.com/amazon-science/chronos-forecasting}{Chronos Github} \\
\hline
LagLlama & Transformer-based & Foundation Model (F) & \href{https://github.com/time-series-foundation-models/lag-llama}{LagLlama Github} \\
\hline
TimesFM & Transformer-based & Foundation Model (F) & \href{https://github.com/google-research/timesfm}{TimesFM Github} \\
\hline
Moirai & Transformer-based & Foundation Model (F) & \href{https://github.com/SalesforceAIResearch/uni2ts}{Moirai Github} \\
\hline
Time-MoE & Transformer-based & Foundation Model (F) & \href{https://github.com/Time-MoE/Time-MoE}{Time-MoE Github} \\
\hline
Sundial & Transformer-based & Foundation Model (F) & \href{https://github.com/thuml/Sundial}{Sundial Github} \\
\hline
PatchTST & Transformer-based & Trained-from-Scratch (TFS) & \href{https://github.com/Nixtla/statsforecast}{NeuralForecast} \\
\hline
Vanilla Transformer & Transformer-based & Trained-from-Scratch (TFS) & \href{https://github.com/Nixtla/statsforecast}{NeuralForecast} \\
\hline
iTransformer & Transformer-based & Trained-from-Scratch (TFS) & \href{https://github.com/Nixtla/statsforecast}{NeuralForecast} \\
\hline
Temporal Fusion Transformer & LSTM with Attention & Trained-from-Scratch (TFS) & \href{https://github.com/Nixtla/statsforecast}{NeuralForecast} \\
\hline
Seasonal Average & Statistical & Baseline (B) & \href{https://pandas.pydata.org/}{Custom with Pandas} \\
\hline
\hline
\end{tabular}%
\label{tab:model_architectures}
\end{table*}

\subsection{Training and Evaluation}
We follow the approach of a time series cross-validation \cite{Bergmeir.2012cross-validation} with a calibration window. The TFS Transformers are trained on a fixed window size (365 days) of the most recent observations up to the split date. After the initial forecast, both the split date and the calibration window are moved forward by the duration of the horizon. Consequently, the training data now encompasses the most recent 365 days, starting from this updated split date. The TFS models are retrained from scratch on the updated calibration window, generating new forecasts for the next horizon period.

This process is repeated until the end of the data is reached. Multiple time series start only during the test period. They will also be picked up during the evaluation steps when the time series length is sufficient, at least greater than the input size and horizon.

Following the insights that the global is the best training approach \cite{Hertel.2023}, we train the TFS Transformers on all the datasets' training splits together.

We ensure that the models only use the intended input size for their predictions by consistently cutting the data to the appropriate input size before the models make their predictions. For example, an input size of 24 defines that the models use the last 24 hours for the forecast. This applies to all tested models: foundation models, TFS Transformers and the Seasonal Average.

For technical reasons, not all models were able to generate forecasts for every scenario. Therefore, we restricted the analysis to the subset of data for which all models provided forecasts. The resulting data reduction was only 0.2\%. In total each models will be evaluated with the different setups on over 6 million forecasting points.

\subsection{Challenges in Evaluating Foundation Models}
While we prevent information leakage of global temporal patterns in the globally trained TFS Transformer models, there remains a special challenge in evaluating time series foundation models like Chronos \cite{Ansari.2024}, TimesFM\cite{Das2024} or LagLlama \cite{Rasul.2024}. The pre-trained models might have learned global temporal patterns (e.g., Covid-19 \cite{Prabowo2023}, geopolitical crises) due to the massive amount of data used during training. Any evaluation on a dataset that is in the same date range as the training data of the foundation model could be potentially affected by such information leakage. Even cross-domain influences are possible, e.g., as weather affects household energy consumption \cite{kang2022weatherelectricityconsumption}, a foundation model trained on historical weather data could theoretically lead to information leakage on a household energy dataset in the same time period.

None of the original foundation model papers address this potential problem. Whether global temporal patterns significantly impact foundation model performance is open to research and outside the scope of this paper. The only possible solution to this problem would be an evaluation based on new data collected after the foundation models were trained.

\subsection{Metrics}
\label{section:metrics}
The prediction results and the ground truth are normalized by subtracting the means and dividing by the standard deviation to be able to compare the time series with different demand ranges, for instance, given by the sizes of the houses and the number of occupants. The focus of the evaluation is the relative performance between models.

The main metric used for the evaluation is shown in Equation \ref{eq:maeh}. It is a slight variation of a traditional Mean Absolute Error (MAE) by averaging the mean absolute errors across categories, which we call \textit{MAE\textsubscript{h}} (MAE households). The metric helps to mitigate the impact of outliers or extreme values within any single category. Also, it diminishes the impact that a house with more observations in the test set has on the final evaluation. This is important to avoid bias since there are large differences in the lengths of the time series in the datasets.

\begin{equation}
\mathrm{MAE}_{h} = \frac{1}{h} \sum_{h=1}^{h} \left( \frac{1}{n} \sum_{i=1}^{n} | y_{h,i} - \hat{y}_{h,i}| \right) 
\label{eq:maeh}
\end{equation}

Where:
\begin{itemize}
    \item $\hat{y}_{h,i}$ is the prediction for household $h$ for prediction $i$ out of $n$
    \item $y_{h,i}$ is the actual value for household $h$ for prediction $i$ of $n$
    \item $h$ is the number of households
    \item $n$ is the number of predictions per household
\end{itemize}

The same logic is applied to compute the Mean Squared Error per household (\textit{MSE\textsubscript{h}}). In addition, we use the adjusted p-norm error per house (\textit{APNE\textsubscript{h}}), introduced by \citet{haben2014new}, which is specifically designed to address the “double penalty” effect. This effect occurs when a forecast that correctly predicts the magnitude of the target (such as a peak) but is slightly displaced in time and is penalized more heavily than a constant, less informative forecast. Traditional point-wise metrics like MAE or MSE fail to account for such temporal misalignments. The adjusted p-norm error mitigates this by searching for a restricted temporal permutation of the forecast that minimizes the error according to a specified p-norm. This search is constrained by an adjustment limit window w, which defines the maximum allowable shift between the forecasted and actual time points. Following the recommendation of the authors, we use a p-norm of 4 and only adapt the adjustment window from w = 3 to w = 1 as we have hourly data instead of half-hourly data.

Based on the various occurrences of zero or near-zero values, we discard other metrics, such as the mean percentage error (MAPE) and symmetric MAPE (SMAPE), as they tend to increase drastically with values near to zero \cite{Hyndman2006}.

To compare predictive accuracy between two forecasting models, we use the Diebold-Mariano (DM) test, which evaluates the null of equal expected loss by testing whether the mean loss differential $d_t = L(e_{1t})-L(e_{2t})$ equals zero \cite{diebold_comparing_1995}.
Forecasts are generated on a non-overlapping multi-step schedule (e.g., one 24-hour-ahead error per 24-hour block). We apply the DM test on these aggregated block losses assuming independence between non-overlapping blocks $(h=1)$ \cite{diebold_comparing_1995}. We use both $\alpha = 0.05$ and $\alpha = 0.01$ for the test statistics.

To ensure better comparability and to avoid imbalances between datasets, a separate ranking was established for each household across all datasets based on the (\textit{MAE\textsubscript{h}}) metric. Afterwards, an average ranking was calculated across all these household rankings. This approach allows for fairer model comparisons, especially since three models could not be evaluated on the Ideal dataset and therefore had to be excluded from the analysis of that particular dataset.

\begin{table*}[ht!]
\begin{center}
\captionof{table}{Model’s \textit{MAE\textsubscript{h}}, \textit{MSE\textsubscript{h}} and \textit{APNE\textsubscript{h}} scores across all datasets}
\scriptsize
\begin{tabular}{llllllllllllll}
\hline \hline
& &  \multicolumn{3}{c}{IDEAL}  &  \multicolumn{3}{c}{Lower Saxony}& \multicolumn{3}{c}{REFIT}  & \multicolumn{3}{c}{Southern Germany} \\
Type & Model & \textit{MAE\textsubscript{h}} & \textit{MSE\textsubscript{h}} & \textit{APNE\textsubscript{h}} & \textit{MAE\textsubscript{h}} & \textit{MSE\textsubscript{h}} & \textit{APNE\textsubscript{h}} & \textit{MAE\textsubscript{h}} & \textit{MSE\textsubscript{h}} & \textit{APNE\textsubscript{h}} & \textit{MAE\textsubscript{h}} & \textit{MSE\textsubscript{h}} & \textit{APNE\textsubscript{h}}  \\
\hline \hline
\multirow{8}{*}{TSFM} & Chronos & 0.537 & 1.184 & 2.194 & 0.528 & 1.167 & 2.211 & 0.576 & 1.123 & 2.105 & 0.521 & 0.923 & 1.778 \\
 & Chronos-Bolt & \underline{0.513} & 1.010 & 2.073 & \textbf{0.490} & 0.931 & 2.019 & \textbf{0.528} & 0.880 & 1.932 & \textbf{0.478} & 0.731 & 1.616 \\
 & LagLLama & 0.647 & 1.366 & 2.269 & 0.670 & 1.388 & 2.275 & 0.743 & 1.445 & 2.229 & 0.646 & 1.153 & 1.876 \\
 & Moirai 1.1 & -* & -* & -* & 1.080 & 3.305 & 4.740 & 1.768 & $>$10 & $>$10 & 1.149 & 4.957 & 6.360 \\
 & Sundial & -* & -* & -* & 0.499 & \underline{0.894} & \underline{1.986} & \underline{0.529} & \textbf{0.827} & \underline{1.888} & 0.497 & \underline{0.714} & \underline{1.594} \\
 & Time-MoE & -* & -* & -* & 0.533 & \textbf{0.892} & \textbf{1.958} & 0.566 & \underline{0.845} & \textbf{1.881} & 0.521 & \textbf{0.710} & \textbf{1.563} \\
 & TimesFM & 0.514 & \underline{0.998} & \underline{2.063} & \underline{0.491} & 0.929 & 2.026 & 0.530 & 0.885 & 1.939 & \underline{0.478} & 0.722 & 1.627 \\
 & TimesFM 2.0 & \textbf{0.509} & 1.035 & 2.102 & 0.493 & 0.974 & 2.069 & 0.532 & 0.919 & 1.976 & 0.485 & 0.758 & 1.654 \\
\hline
\multirow{4}{*}{TFS} & PatchTST & 0.516 & \textbf{0.969} & \textbf{2.048} & 0.494 & 0.955 & 2.058 & 0.535 & 0.899 & 1.960 & 0.499 & 0.760 & 1.652 \\
 & TFT & 0.577 & 1.126 & 2.144 & 0.580 & 1.147 & 2.161 & 0.635 & 1.171 & 2.117 & 0.616 & 0.973 & 1.778 \\
 & VanillaTransformer & 0.549 & 1.075 & 2.124 & 0.556 & 1.098 & 2.141 & 0.603 & 1.113 & 2.097 & 0.573 & 0.896 & 1.744 \\
 & iTransformer & 0.589 & 1.133 & 2.143 & 0.588 & 1.103 & 2.123 & 0.648 & 1.117 & 2.068 & 0.598 & 0.922 & 1.737 \\
\hline
\multirow{2}{*}{Baseline} & Naive-Forecast & 0.703 & 1.502 & 2.217 & 0.608 & 1.311 & 2.266 & 0.733 & 1.280 & 2.060 & 0.718 & 1.272 & 1.868 \\
 & SeasonalAverage & 0.609 & 1.208 & 2.128 & 0.572 & 1.102 & 2.071 & 0.577 & 0.982 & 1.965 & 0.539 & 0.854 & 1.696 \\
\hline \hline
\end{tabular}
\newline
\scriptsize $^{*}$Dataset included in TSFM pre-training data.
\label{evaluation_overall}
\end{center}
\end{table*}
\section{Results} \label{sec:results}

Table \ref{evaluation_overall} shows the \textit{MAE\textsubscript{h}}, \textit{MSE\textsubscript{h}} and \textit{APNE\textsubscript{h}} for all models across the different datasets. Comparing the different metrics, there is no single model that dominates the benchmark. Stepping back, it can be seen that most TSFMs, though not all, outperform the other approaches. Except for the IDEAL dataset, all best or second-best results are achieved by TSFM.
The results of the Diebold-Mariano test confirm that statistically significant performance differences exist in 91.8\% of the pairwise comparisons across all models, horizons, and input sizes ($p < 0.05$). In only 8.2\% of the cases, no statistically significant difference was observed, suggesting comparable model performance or context-dependent advantages. This trend remains robust even under a stricter significance level of $\alpha = 0.01$, where only 9.1\% of the pairwise comparisons yield non-significant results.
A major contributor to these non-significant results is the \textit{Moirai} model, which accounts for over half of the instances where no clear statistical winner could be determined. This indicates that while \textit{Moirai} delivers competitive forecasts in certain scenarios, it fails to consistently outperform other models across the benchmarked tasks.
Other examples of such non-significant pairs include mostly models with very close performances like Chronos-Bolt vs.\ TimesFM 2.0 (horizon 24h, input size 96h), iTransformer vs.\ TFT (horizon 168h, input size 168h), and Chronos vs.\ VanillaTransformer (horizon 168h, input size 24h). Especially when comparing good performing models like Chronos-Bolt, TimesFM oder Sundial with TFS models, the performance comparisons are always significant starting with an input size of 96. Consequently, the performance rankings established in this benchmark are statistically robust and not driven by random variance.

For the \textit{MAE\textsubscript{h}} Chronos-Bolt delivers best performances across the Lower Saxony, REFIT and Southern Germany datasets, often with TimesFM following closely. The newer version TimesFM 2.0 outperformed Chronos-Bolt on the IDEAL dataset.
But also the TFS Transformer PatchTST is performing good or even outperforming the other models on the IDEAL dataset in terms of \textit{MSE\textsubscript{h}}. On the other datasets Time-MoE and Sundial have the best performance for the \textit{MSE\textsubscript{h}} metric. The worst-performing models overall were Lag-LLama and Moirai, which could not outperform the SeasonalAverage baseline.

In terms of \textit{APNE\textsubscript{h}}, which lies the focus on load peak prediction, Time-MoE and Sundial are slightly outperforming the other models, indicating they try more to predict load peaks, but seem to be sometimes off by a timestep. In general, the gap between TSFM and the baseline SeasonalAverage is the smallest on the \textit{APNE\textsubscript{h}} metric.

A more detailed analysis how the models behave with different input sizes and horizons allows Table \ref{evaluation_all_datasets}, which shows the average rank based on \textit{MAE\textsubscript{h}} and the \textit{MAE\textsubscript{h}} by model, input size, and horizon.

Also, in this case there is no dominant model, but a clear pattern is visible: With a short input size, the TFS Transformer PatchTST is the best model but looses its position against the TSFM when the input size increases.
Especially Sundial has a slight advantage over other TSFM with input sizes longer than 24 hours. Overall, the TSFM models and also PatchTST perform better when provided with a longer input size but also the advantage of all models against the baseline decreases.

Furthermore, the models' mean error increases with longer horizons, while the performance loss remains most of the time limited. The SeasonalAverage's performance also increases with the input size, making it a strong baseline. An exception is the VanillaTransformer which performance remains the same.

TimesFM 2.0 frequently achieves one of the lowest ranks among all models, indicating that it performs very well on many time series. However, it struggles with certain cases, which leads to a slight disadvantage in the overall \textit{MAE\textsubscript{h}}.

\begin{table*}[h]
\centering
\renewcommand{\arraystretch}{0.65}
\caption{Rank (MAE\textsubscript{h}) and MAE\textsubscript{h} results for every input size and horizon for all datasets. Best results are in bold, second best underlined.}
\scriptsize
\begin{tabular}{lllrrrrrr}
\hline \hline
Type & Model & Input & \multicolumn{6}{c}{Horizon} \\
 &  & size &  \multicolumn{2}{c}{24} &  \multicolumn{2}{c}{96} &  \multicolumn{2}{c}{168} \\
 & & & MAE\textsubscript{h} & RANK & MAE\textsubscript{h} & RANK & MAE\textsubscript{h} & RANK \\
\hline \hline
\multirow{8}{*}{TSFM} & Chronos & 24 & 0.55 & 5.29 & 0.55 & 4.82 & 0.57 & 4.63 \\
 & Chronos-Bolt & 24 & 0.53 & 4.13 & 0.54 & 3.94 & 0.56 & 4.10 \\
 & LagLLama & 24 & 0.60 & 8.59 & 0.78 & 10.86 & 0.82 & 10.77 \\
 & Moirai 1.1* & 24 & 0.75 & 13.50 & 1.98 & 14.00 & 2.56 & 14.00 \\
 & Sundial* & 24 & 0.53 & 5.17 & 0.56 & 5.23 & 0.56 & 4.88 \\
 & Time-MoE* & 24 & 0.59 & 9.40 & 0.58 & 7.83 & 0.58 & 7.04 \\
 & TimesFM & 24 & \underline{0.53} & \underline{3.76} & 0.54 & 3.65 & 0.57 & 5.17 \\
 & TimesFM 2.0 & 24 & 0.53 & 4.16 & \underline{0.54} & \underline{3.52} & \underline{0.55} & \underline{3.34} \\
\hline
\multirow{4}{*}{TFS} & PatchTST & 24 & \textbf{0.51} & \textbf{2.36} & \textbf{0.53} & \textbf{2.53} & \textbf{0.54} & \textbf{2.65} \\
 & TFT & 24 & 0.56 & 6.76 & 0.58 & 6.74 & 0.59 & 6.69 \\
 & VanillaTransformer & 24 & 0.53 & 3.81 & 0.56 & 5.03 & 0.57 & 5.07 \\
 & iTransformer & 24 & 0.62 & 9.69 & 0.60 & 8.26 & 0.60 & 7.40 \\
\hline
\multirow{2}{*}{Baseline} & Naive-Forecast & 24 & 0.68 & 9.77 & 0.69 & 9.18 & 0.71 & 9.00 \\
 & SeasonalAverage & 24 & 0.64 & 9.73 & 0.67 & 9.70 & 0.71 & 9.62 \\
\hline\hline
\multirow{8}{*}{TSFM} & Chronos & 96 & 0.52 & 5.11 & 0.54 & 4.89 & 0.55 & 5.50 \\
 & Chronos-Bolt & 96 & 0.49 & 2.95 & 0.51 & \underline{2.86} & \underline{0.50} & \textbf{2.44} \\
 & LagLLama & 96 & 0.58 & 8.58 & 0.66 & 10.19 & 0.68 & 10.45 \\
 & Moirai 1.1* & 96 & 0.79 & 12.10 & 0.88 & 13.81 & 1.70 & 14.00 \\
 & Sundial* & 96 & \textbf{0.48} & 3.27 & \textbf{0.50} & 3.37 & \textbf{0.50} & 3.25 \\
 & Time-MoE* & 96 & 0.53 & 7.96 & 0.53 & 6.46 & 0.53 & 6.88 \\
 & TimesFM & 96 & 0.49 & \underline{2.86} & 0.50 & 2.89 & 0.51 & 3.59 \\
 & TimesFM 2.0 & 96 & \underline{0.49} & \textbf{2.52} & \underline{0.50} & \textbf{2.50} & 0.51 & \underline{2.95} \\
\hline
\multirow{4}{*}{TFS} & PatchTST & 96 & 0.50 & 3.88 & 0.51 & 3.77 & 0.52 & 3.73 \\
 & TFT & 96 & 0.58 & 8.61 & 0.59 & 8.44 & 0.57 & 7.27 \\
 & VanillaTransformer & 96 & 0.53 & 5.74 & 0.57 & 6.82 & 0.56 & 7.00 \\
 & iTransformer & 96 & 0.61 & 10.06 & 0.59 & 8.56 & 0.57 & 7.94 \\
\hline
\multirow{2}{*}{Baseline} & Naive-Forecast & 96 & 0.68 & 10.19 & 0.69 & 9.80 & 0.71 & 10.05 \\
 & SeasonalAverage & 96 & 0.57 & 8.43 & 0.58 & 8.14 & 0.58 & 7.88 \\
\hline\hline
\multirow{8}{*}{TSFM} & Chronos & 168 & 0.51 & 4.92 & 0.52 & 5.08 & 0.54 & 5.34 \\
 & Chronos-Bolt & 168 & \underline{0.47} & \textbf{2.12} & 0.49 & \underline{2.62} & \underline{0.49} & \textbf{2.05} \\
 & LagLLama & 168 & 0.55 & 7.91 & 0.60 & 9.65 & 0.62 & 9.82 \\
 & Moirai 1.1* & 168 & 1.12 & 11.10 & 1.20 & 11.50 & 0.84 & 13.60 \\
 & Sundial* & 168 & \textbf{0.47} & 3.54 & \textbf{0.49} & 3.50 & \textbf{0.49} & 3.54 \\
 & Time-MoE* & 168 & 0.51 & 7.12 & 0.51 & 6.13 & 0.52 & 6.38 \\
 & TimesFM & 168 & 0.48 & 2.63 & 0.49 & 2.85 & 0.50 & 2.89 \\
 & TimesFM 2.0 & 168 & 0.47 & \underline{2.22} & \underline{0.49} & \textbf{2.18} & 0.49 & \underline{2.60} \\
\hline
\multirow{4}{*}{TFS} & PatchTST & 168 & 0.50 & 4.76 & 0.51 & 4.39 & 0.51 & 4.67 \\
 & TFT & 168 & 0.58 & 9.28 & 0.60 & 9.14 & 0.59 & 8.34 \\
 & VanillaTransformer & 168 & 0.53 & 6.73 & 0.57 & 7.41 & 0.57 & 7.17 \\
 & iTransformer & 168 & 0.61 & 10.47 & 0.57 & 8.16 & 0.58 & 8.29 \\
\hline
\multirow{2}{*}{Baseline} & Naive-Forecast & 168 & 0.68 & 10.42 & 0.69 & 10.29 & 0.71 & 10.22 \\
 & SeasonalAverage & 168 & 0.55 & 7.74 & 0.55 & 7.57 & 0.56 & 7.53 \\
\hline \hline
\end{tabular}
\newline
\scriptsize $^{*}$IDEAL dataset excluded.
\label{evaluation_all_datasets}
\end{table*}
\section{Discussion}

Our experiments on four datasets, namely Lower Saxony, Southern Germany, IDEAL, and REFIT, 
show that while the best TSFM, like TimesFM, Chronos-Bolt, Time-MoE, and Sundial outperform the best TFS Transformer, like PatchTST, not all TSFM perform equally well. This directly addresses the stated research question, ``Can zero-shot TSFMs match the capabilities of state-of-the-art trained-from-scratch Transformers in forecasting household electricity load?''. Our empirical evaluation demonstrates that zero-shot TSFMs not only achieve performance on par with TFS Transformers but, in certain cases, even surpass them in the context of household electricity STLF.

Going more into detail and considering the input sizes, TSFM like Sundial, TimesFM or Chronos-Bolt significantly outperform PatchTST for longer input sizes (96 and 168) in regards of MAE. This suggests that the performance of foundation models may increase with input size. A possible reason is that TSFM need more context than custom-trained Transformers to identify the patterns of the time series. 

LagLlama performs poorly for all input sizes compared to the other foundation models and most TFS Transformers. This might be explained by LagLlama's architecture, which uses lags to predict future values \cite{Rasul.2024}. These lags include quarterly, monthly, weekly, daily, and hourly levels \cite{Rasul.2024}, while our defined maximum input size of 168 hours allows only incorporating daily to weekly lags. Similarly our restricted experimental setup with limited information about the time series could also explain the performance of the Moirai model, which was originally evaluated with a significantly longer context size of 1000 \cite{woo_unified_2024}.

Considering the performance of the TFS Transformer PatchTST, our study supports the findings of \citet{Hertel.2023} and \citet{Cen.2024}, as the PatchTST model provides good overall performance and even slightly outperforms other TSFM on an input size of 24 hours. Additionally, the results of \citet{Hertel.2023} indicate that TFS Transformers perform better when trained globally with more data. Our study extends their findings by showing that the zero-shot performance of the also globally trained TSFM is better or comparable to TFS Transformers. Interestingly, we could not reproduce the statement that input sizes exceeding the horizon decreases forecasting performance of transformer models \cite{wen_Transformers_survery_2022}, but observed the opposite: PatchTST as well as all TSFM benefit from longer context. Just the the VanillaTransformer did not benefit from longer input sizes matching the results of \cite{wen_Transformers_survery_2022}.

Moreover, when assessed using the household STLF-specific metric (\textit{APNE\textsubscript{h}}), TSFM models like Time-MoE and Sundial are able to deliver good results, but the difference to the baseline is significantly smaller. Most TSFM seem to predict more conservative without extreme load peaks. As these peaks are also relevant in STLF, e.g. regarding grid capacity and stability, TSFM seem not to be suitable for predicting these load peaks.

A final observation of our analysis is that the SeasonalAverage baseline outperforms some of the TFS Transformers and TSFM for horizons of 96 hours and 168 hours. This is probably due to the strong daily patterns present in household energy load, which makes the SeasonalAverage a suitable baseline.

Compared to standard TSFM evaluations (see Section \ref{section:related-work}), our approach combines multi-dataset evaluation with time-series cross-validation for a more comprehensive assessment. This design mitigates the limitations of smaller datasets by capturing both cross-series variation and performance changes over time, thereby strengthening the validity of our findings and supporting stronger claims about the robustness and generalizability of TSFMs relative to the TFS Transformer.

\section{Limitations \& Outlook}
Naturally, our analysis is not without limitations. Hyperparameter tuning the TFS Transformers may increase the performance of these models supporting the findings of \citet{Sievers.2023} and \citet{Upadhyay.2023} while simultaneously increasing the training effort compared to TSFM even more. On the other hand, longer input sizes could lead to better performance of the foundation models, especially for LagLlama and Moirai, which might need higher context lengths due to their models architecture \cite{Rasul.2024, woo_unified_2024}.

There are several directions for future research on household STLF. First, incorporating longer input sizes would allow for drawing a better picture of TFS Transformer-based and foundation model behavior, e.g., the behavior of LagLlama on longer inputs. While TSFM provide zero-shot forecasts that can outperform SOTA TFS Transformers, fine-tuning foundation models has been shown to increase performance in other disciplines, such as foundation LLMs. Moreover TSFM, which have been trained on some household energy data, performed better than other TSFMs. Hence, pre-training and fine-tuning foundation models on energy time series forecasting could further increase their capabilities. Second, we propose that the models can learn cross-domain global patterns, such as the COVID-19 pandemic or geopolitical crises, which may lead to information leakage when evaluating these models on holdout datasets originating from the same time period as the training datasets. This possible problem should be explored in future research. Third, our univariate analysis could be extended to multivariate TSFMs, including covariates such as weather data \cite{kang2022weatherelectricityconsumption}.

\section{Conclusion}
Motivated by recent algorithmic developments in time series forecasting, this study investigated whether time series foundation models are competitive to SOTA TFS Transformers on household STLF tasks.
Following the guideline of \citet{Haben.2021} we considered time series cross-validation on multiple datasets using statistical baselines and sophisticated TFS Transformers. In our benchmark, unlike the TFS Transformers, the TSFMs were used out-of-the-box for prediction without task-specific adaptation or fine-tuning.
We show that TSFM are already capable of delivering competitive and in most cases better forecast performance compared to trained-from-scratch Transformers. The foundation models show their strength, especially when there is more context (i.e., longer input sizes). Furthermore, our findings indicate that, in the case of LagLLama or Moirai, its special architecture may harm performance when dealing with a limited input context.
In conclusion, the ability of foundation models to achieve high accuracy with limited data and without training opens up new possibilities for developing more efficient and accessible energy forecasting solutions.

\bibliographystyle{IEEEtranN} 
\bibliography{refs}

\begin{IEEEbiography}[{\includegraphics[width=1in,height=1.25in,clip,keepaspectratio]{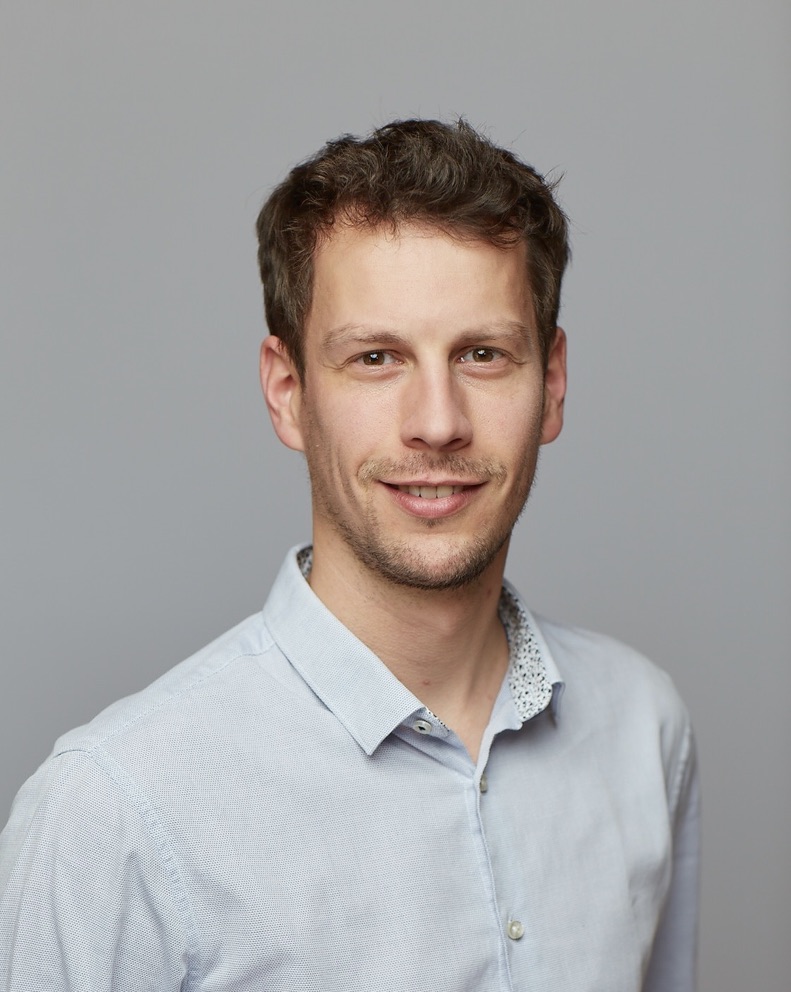}}]{Marcel Meyer} received the B.Sc. and Dipl.-Wi.-Ing. (equivalent M.Sc.) degrees in industrial engineering and management from the Technische Universität Dresden, Germany, in 2017.

From 2018 to 2024, he worked in several positions in the industry, reaching from textile and aerospace industry to data science consultancy. Since 2024, he is Research Assistant at Paderborn University with a focus on digital twins and Time Series Foundation Models.

\end{IEEEbiography}

\begin{IEEEbiography}[{\includegraphics[width=1in,height=1.25in,clip,keepaspectratio]{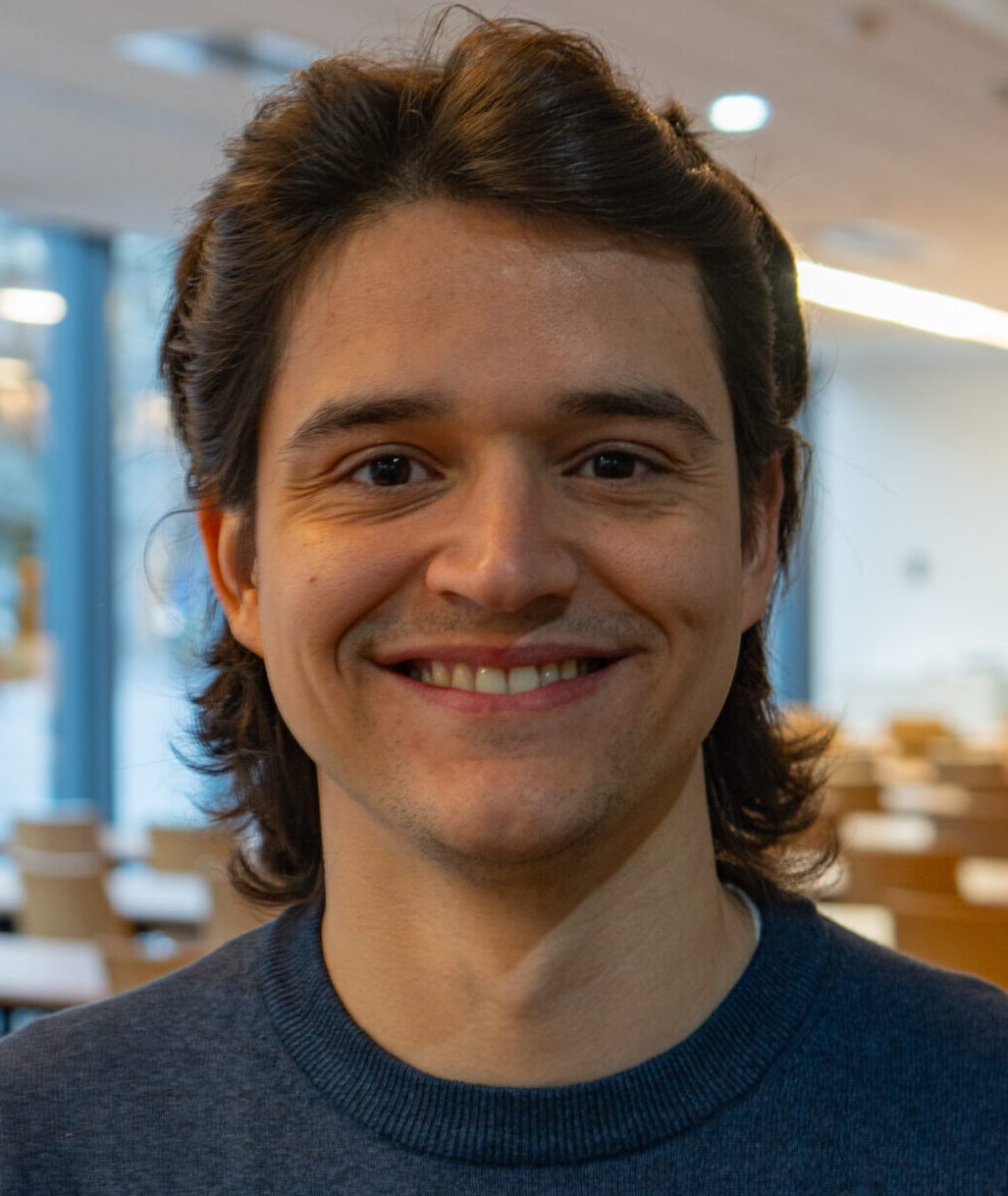}}]{David Zapata González} holds a B.Sc. in Industrial Engineering from Yacambú University (Venezuela, 2017) and a Master’s degree in Production Engineering and Management from the OWL University of Applied Sciences and Arts (Germany, 2022). From 2022 to 2024, he worked as a Data Scientist in Operations at a home appliance manufacturer. He is currently a Research Assistant at Paderborn University, focusing on data-driven modeling of physical systems and industrial processes, with a particular emphasis on energy-related applications.

\end{IEEEbiography}

\begin{IEEEbiography}[{\includegraphics[width=1in,height=1.25in,clip,keepaspectratio]{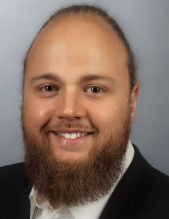}}]{Sascha Kaltenpoth} received the B.Sc. and M.Sc. degrees in business informatics from Paderborn University, in 2021 and 2023, respectively. He is currently a Research Assistant with Paderborn University. His research interests include data science, with a focus on large language models (LLMs) based assistance systems and LLM-based time series forecasting
\end{IEEEbiography}

\begin{IEEEbiography}[{\includegraphics[width=1in,height=1.25in,clip,keepaspectratio]{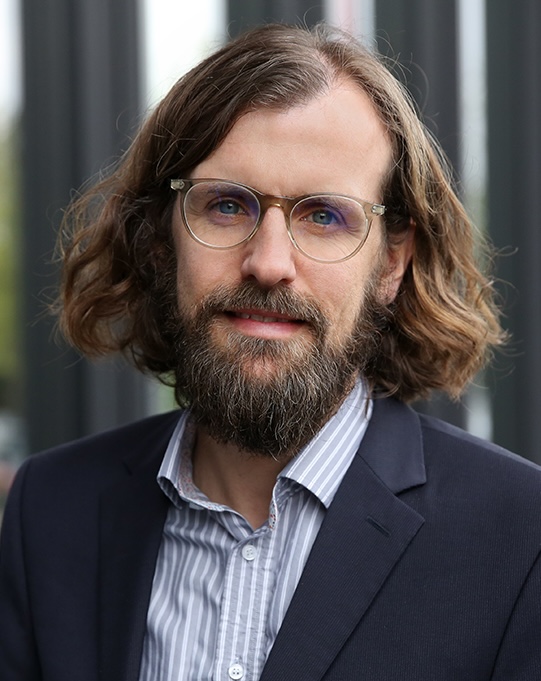}}]{Oliver Müller} received the B.Sc., M.Sc., and Ph.D. degrees in information systems from the School of Business and Economics, University of Münster. He is currently a Professor of management information systems and data analytics with Paderborn University. His research interests include data-driven judgment and decision-making. This includes the design and use of machine learning solutions for supporting human judgment and decision-making, and studying the acceptance and implications of data-driven decision-making in organizations.
\end{IEEEbiography}

\EOD

\end{document}